\documentclass[aps,preprint,showpacs,showkeys]{revtex4}
\usepackage{amsmath}

\usepackage{bm}

\begin{document}

\title[]{Spectra of Interacting Electrons in a Quantum Dot: Quasi-Exact
Solution}
\author{Ramazan Ko\c{c}}
\author{Hayriye T\"{u}t\"{u}nc\"{u}ler}
\author{Eser Ol\u{g}ar}
\email{koc@gantep.edu.tr,tutunculer@gantep.edu.tr,olgar@gantep.edu.tr}
\affiliation{Department of Physics, Faculty of Engineering\\
University of Gaziantep, 27310 Gaziantep Turkey}
\date{[]Received May 3 2004}

\begin{abstract}
We present a procedure to solve the Schr\"{o}dinger equation of two
interacting electrons in a quantum dot in the presence of an external
magnetic field within the context of quasi-exactly-solvable spectral
problems. We show that the symmetries of the Hamiltonian can be recovered
for specific values of the magnetic field, which leads to an exact
determination of the eigenvalues and eigenfunctions. We show that the
problem possesses a hidden $sl_{2}-$algebraic structure.
\end{abstract}

\maketitle

\setcounter{page}{1}

\section{INTRODUCTION}

Recent developments in micro fabrication technology have allowed the
creation of quantum dots (QDs) in semiconductor structures\cite{jac}. The
QDs may be regarded as artificial atoms, with disk-like shapes, in which
electrons are confined by an artificial electrostatic potential instead of
being attracted to a nucleus. The electron confinement within the QD is
described by a Hamiltonian that includes the Coulumb potential, as well as a
parabolic potential. Since no general analytical solutions are available,
the Hamiltonian has been solved numerically\cite{que} by means of the
shifted $1/N$ expansion\cite{said}, the WKB treatment\cite{gar,stan} and the
non-local potential method\cite{alber,lopez}. A particular analytical
solution of the Hamiltonian has been obtained by Taut\cite{taut}. The
renormalization perturbation series method has been proposed by Matulis and
Peeters to calculate the energy spectrum of the quantum dot\cite{mat}, and
exact solutions for two electrons in a parabolic quantum dot have been
obtained by using a power series method\cite{zhu}.

Exact solvability of the Schr\"{o}dinger equation has been the main interest
since the early days of quantum mechanics\cite{levai}. It has been solved
exactly for a large number of potentials by employing various techniques. In
quantum mechanics, there exist potentials for which it is possible to find a
finite number of eigenvalues and associated eigenfunction exactly in closed
form. These systems are said to be quasi-exactly solvable\cite{turb0,bender}%
. Recently, Turbiner pointed out that the Coulomb correlation problem for a
system of electrons in an external oscillator potential is one of the
quasi-exactly-solvable problems\cite{turb}. Hidden symmetries of
two-electron quantum dots have been analyzed, and it has been shown that
under certain conditions, the motion becomes integrable\cite{simon}. This
indicates the existence of symmetries in the spectrum of the Hamiltonian. In
this paper, we investigate the quasi-exact solvability of the interacting
electrons confined in a QD by a parabolic potential in an external magnetic
field. The results obtained here should be useful for checking and assessing
numerical and approximate methods.

In this paper we will give a different treatment of interacting electrons in
a QD. In Section II, we briefly review the construction of the Hamiltonian
of two electrons confined in a QD in the presence of an external magnetic
field. In Section III, we present an algebraic framework to show that the
corresponding Hamiltonian possesses a hidden $sl_{2}-$algebraic structure
and we obtain somewhat simpler recurrence relations, leading to exact
eigenvalues of the Hamiltonian. In Section IV, our results are summarized,
and finally, Section V consists of discussions and concluding remarks.

\section{THEORY}

Consider a QD containing two electrons in the presence of a magnetic field $B
$ perpendicular to the dot and a lateral potential $m^{\ast }\omega
_{0}^{2}r^{2}/2$ that confines the electrons. The Hamiltonian of the present
system is

\begin{equation}
H=\sum\limits_{i=1}^{2}(\frac{1}{2m_{i}^{\ast }}(P_{i}+eA(r_{i}))^{2}+\frac{1%
}{2}m_{i}^{\ast }\omega _{0}^{2}r_{i}^{2})+\frac{e^{2}}{\varepsilon \left|
r_{2}-r_{1}\right| }  \label{e1}
\end{equation}%
where $m^{\ast },\omega _{0},$ and $\varepsilon $ are the effective mass of
the electrons, the confinement frequency and the dielectric constant of the
medium, respectively. $A(r_{i})$ denotes the vector potential at the $i^{th}$
particle's location, and the choice of the symmetric gauge vector potential $%
A(r_{i})=B/2(-y_{i},x_{i},0)$ leads to the following Hamiltonian:

\begin{equation}
H=\sum\limits_{i=1}^{2}(\frac{P_{i}^{2}}{2m_{i}^{\ast }}+\frac{1}{2}%
m_{i}^{\ast }\omega ^{2}r_{i}^{2})+\frac{e^{2}}{\epsilon \left|
r_{2}-r_{1}\right| }+\frac{1}{2}\omega _{c}L_{tot},  \label{e2}
\end{equation}%
where $\omega _{c}=eB/m^{\ast }$ stands for the cyclotron frequency of the
electron, $\omega =\sqrt{\omega _{0}^{2}+\left( \frac{\omega _{c}}{2}\right)
^{2}}$ is the effective frequency, and $L_{tot}$ is the total angular
momentum along the $z$-direction. In order to show that the Hamiltonian in
Eq. (\ref{e2}) is separable, we just need a straightforward analysis.

Now we introduce relative and center-of-mass coordinates:

\begin{equation}
r=r_{2}-r_{1},\quad R=\frac{1}{2}(r_{1}+r_{2}).  \label{e3}
\end{equation}%
In the new coordinates, the Hamiltonian in Eq. (\ref{e2}) is then separated
into two parts, and it can be written as%
\begin{equation}
H=2H_{r}+\frac{1}{2}H_{R},  \label{e4}
\end{equation}%
where 
\begin{subequations}
\begin{eqnarray}
H_{r} &=&\frac{p^{2}}{2m^{\ast }}+\frac{1}{2}m^{\ast }\omega ^{2}r^{2}+\frac{%
e^{2}}{2\epsilon r}+\frac{1}{2}\omega _{c}L_{r},  \label{e5a} \\
H_{R} &=&\frac{P^{2}}{2m^{\ast }}+\frac{1}{2}m^{\ast }\omega ^{2}R^{2}+\frac{%
1}{2}\omega _{c}L_{R}.  \label{e5b}
\end{eqnarray}%
It is obvious that $L_{tot}=L_{r}+L_{R}$. Equation (\ref{e5b}) is the
Hamiltonian of the harmonic oscillator, and its eigenvalues are given by 
\end{subequations}
\begin{equation}
E_{R}=(2N+\left| M\right| +1)\hbar \omega +M\frac{\hbar \omega _{c}}{2},
\label{e6}
\end{equation}%
where $N=0,1,2$,$\cdots $ denotes the radial quantum number and $M=0,\pm
1,\pm 2,\cdots $ is the azimuthal quantum number. Let us turn our attention
to the solution of the Hamiltonian $H_{r}$. In the polar coordinate $%
r=(r,\alpha )$, if the eigenfunction 
\begin{equation}
\phi =r^{-\frac{1}{2}}e^{im\alpha }u(r)  \label{e7}
\end{equation}%
is introduced, the Schr\"{o}dinger equation $H_{r}\phi =E_{r}\phi $, can be
expressed as%
\begin{equation}
\left( -\frac{\hbar ^{2}}{2m^{\ast }}\frac{d^{2}}{dr^{2}}+\frac{\hbar ^{2}}{%
2m^{\ast }}(m^{2}+\frac{1}{4})\frac{1}{r^{2}}+\frac{1}{2}m^{\ast }\omega
^{2}r^{2}+\frac{e^{2}}{2\epsilon r}+\frac{1}{2}\omega _{c}L_{r}\right)
u(r)=E_{r}u(r)  \label{e8}
\end{equation}%
From now on we restrict ourselves to the solution of \ Eq. (\ref{e8}).

\section{Method of Solution}

In this section we show that the Schr\"{o}dinger equation, Eq. (\ref{e8}),
is one of the recently discovered quasi-exactly solvable operators \cite%
{turb0,bender,turb}. It is well known that the underlying idea behind the
quasi-exact solvability is the existence of a hidden algebraic structure.
Let us introduce the following realization of the $sl_{2}-$algebra: 
\begin{equation}
J_{+}=r^{2}\frac{d}{dr}-jr,\quad J_{-}=\frac{d}{dr},\quad J_{0}=r\frac{d}{dr}%
-\frac{j}{2}.  \label{e9}
\end{equation}%
The generators satisfy the commutation relations of the $sl_{2}-$algebra for
any value of the parameter $j$. If $j$ is a positive integer, the algebra in
Eq. (\ref{e8}) possesses a $j+1-$dimensional irreducible representation: 
\begin{equation}
P_{j+1}=\left\langle 1,r,r^{2},\ldots ,r^{j}\right\rangle .  \label{e10}
\end{equation}%
Linear and bilinear combinations of the operators given in Eq. (\ref{e9})
are quasi-exactly solvable when the space is defined as in Eq. (\ref{e10}).
In order to show that the Schr\"{o}dinger equation has a $sl_{2}$-symmetry,
let us consider the following combinations of the operators in Eq. (\ref{e9}%
): 
\begin{equation}
T=-J_{-}J_{0}-\frac{1}{2}(4m+j)J_{-}+\hbar \omega J_{+}.  \label{e11}
\end{equation}%
Then, the eigenvalue problem becomes 
\begin{equation}
TP_{k}(r)=\lambda P_{k}(r)  \label{e12}
\end{equation}%
where $P_{k}(r)$ is the $k^{th}$ degree polynomial in $r$. After changing
the variable $r\rightarrow \frac{\hbar }{\sqrt{2m^{\ast }}}r$ and
substituting%
\begin{equation}
u(r)=r^{m+\frac{1}{2}}e^{-\frac{\hbar \omega }{4}r^{2}}P_{k}(r),  \label{e13}
\end{equation}%
we can show that the eigenvalue equations, Eqs. (\ref{e8}) and (\ref{e12}),
are identical when the following holds:%
\begin{equation}
E_{r}=(j+\left| m\right| +1)\hbar \omega +\frac{1}{2}m\hbar \omega
_{c},\quad \lambda =-\frac{e^{2}\sqrt{m^{\ast }}}{\sqrt{2}\varepsilon \hbar }%
.  \label{e14}
\end{equation}%
When the generators act on the polynomial in Eq. (\ref{e10}), we can obtain
the following recurrence relation: 
\begin{equation}
\hbar \omega (k-j)P_{k+1}(\lambda )-\lambda P_{k}(\lambda
)-k(k+2m)P_{k-1}(\lambda )=0  \label{e15}
\end{equation}%
with the initial condition $P_{0}(1)=1$. If $\lambda _{i}$ is a root of the
polynomial $P_{k+1}(\lambda )$, the wavefunction is truncated at $k=j$ and
belongs to the spectrum of the Hamiltonian $T$. This property implies that
the wavefunction is itself the generating function of the energy polynomials.

The roots of the recurrence relation in Eq. (\ref{e15}) can be computed, and
the first few of them are given by%
\begin{eqnarray}
P_{1}(\lambda ) &=&\lambda ,  \notag \\
P_{2}(\lambda ) &=&\lambda ^{2}-j(1+2m)\hbar \omega ,  \notag \\
P_{3}(\lambda ) &=&\lambda \lbrack \lambda ^{2}-(5j+6km-4m-4)]\hbar \omega ,
\label{e16} \\
P_{4}(\lambda ) &=&\lambda ^{4}-2\lambda ^{2}[j(6m+7)-8m-11]\hbar \omega
+j(j-2)(1+2m)(3+2m)(\hbar \omega )^{2}.  \notag
\end{eqnarray}%
The function $P_{j}(r)$ forms a basis for $sl_{2}-$algebra, and it can be
written in the form 
\begin{equation}
P_{j}(r)=\sum\limits_{k=0}^{j}\frac{j!(2m)!P_{k}(\lambda )}{(j-k)!k!(2m+k)!}%
(\hbar \omega r)^{n}.  \label{e17}
\end{equation}%
Therefore, we can obtain the eigenfunction of Eq. (\ref{e1}) in a closed
form, at least, the first few values of $j$.

\section{Results}

The polynomial $P_{k}(\lambda )$ vanishes for $k=j+1,$ and the roots of the
polynomials are given by 
\begin{eqnarray}
\lambda  &=&0,  \notag \\
\lambda  &=&\mp \sqrt{(1+2m)\hbar \omega },  \notag \\
\lambda  &=&0,\;\mp \sqrt{(3+4m)\hbar \omega },  \label{e18} \\
\lambda  &=&\mp \sqrt{(10+10m\mp \sqrt{73+128m+64m^{2}})\hbar \omega } 
\notag
\end{eqnarray}%
for $j=0,1,2$ and $3$, respectively. It is obvious that the effective
frequency $\omega $ can be expressed as%
\begin{equation}
\omega =\frac{\lambda ^{2}}{\hbar \eta (j,m)},  \label{e19}
\end{equation}%
where $\eta (j,m)$ is the coefficients of $\hbar \omega $ in \ Eq. (\ref{e18}%
). The relation in Eq. (\ref{e19}) leads to the following conclusion: The
Coulomb interaction destroys the general symmetry of the QD, but the
magnetic field can restore the symmetries which are common for a harmonic
oscillator and Coulomb system. This feature implies that for some specific
values of the magnetic field $B,$ the Hamiltonian in Eq. (\ref{e1}) can be
solved exactly. Since the effective frequency $\omega $ is a function of $B,$
the relation in Eq. (\ref{e19}) can be written as 
\begin{equation}
\omega _{c}\equiv \frac{eB}{m^{\ast }}=\sqrt{\left( \frac{e^{4}m^{\ast }}{%
\eta \epsilon ^{2}\hbar ^{3}}\right) ^{2}-\omega _{0}^{2}}.  \label{e20}
\end{equation}

In order to find the exact eigenvalues and eigenfunctions of Eq. (\ref{e1}),
it is enough to perform the calculation in Eq. (\ref{e20}). The accuracy of
the approximate and the numerical results can be checked by means of the
exact solution given in the previous section. Let us consider the $GaAs$
heterostructure. The material parameters of this structure are chosen to be $%
m^{\ast }=0.067m_{0}$ and $\varepsilon =12.4$. The magnetic field is
calculated and tabulated in Table 1. The confinement energy $\hbar \omega
_{0}$ is given as $4\quad meV$.

\begin{table}[t]
\begin{tabular}{|c|c|c|c|c|c|c|c|}
\hline
& \multicolumn{3}{|c|}{$\hbar \omega _{c}$} & \multicolumn{3}{|c|}{$E_{r}$}
&  \\ \hline
$j$ & $m=0$ & $m=1$ & $m=2$ & $m=0$ & $m=1$ & $m=2$ & $N_{r}$ \\ \hline
$1$ & $1.870092$ & $0.623318$ & $0.373936$ & $1.870109$ & $1.246710$ & $%
1.121980$ & $0$ \\ \hline
$2$ & $0.311582$ & $0.133339$ & $0.084627$ & $0.467527$ & $0.333828$ & $%
0.297140$ & $0$ \\ \hline
$3$ & $0.100529$ & $0.050923$ & $0.033572$ & $0.201694$ & $0.154332$ & $%
0.137109$ & $0$ \\ \hline
$3$ & $1.284394$ & $0.502495$ & $0.321600$ & $2.568837$ & $1.507640$ & $%
1.286700$ & $1$ \\ \hline
$4$ & $0.043551$ & $0.024225$ & $0.015929$ & $0.110700$ & $0.088650$ & $%
0.078318$ & $0$ \\ \hline
$4$ & $0.240664$ & $0.111710$ & $0.074112$ & $0.601993$ & $0.391842$ & $%
0.335014$ & $1$ \\ \hline
$5$ & $0.021771$ & $0.012144$ & $0.006844$ & $0.069584$ & $0.056970$ & $%
0.048959$ & $0$ \\ \hline
$5$ & $0.082677$ & $0.043767$ & $0.029690$ & $0.249192$ & $0.177607$ & $%
0.152686$ & $1$ \\ \hline
$5$ & $1.005801$ & $0.427792$ & $0.284926$ & $3.0174987$ & $1.711430$ & $%
1.425080$ & $2$ \\ \hline
\end{tabular}%
\caption{The cyclotron frequency of the electron and the corresponding
eigenvalues.}
\label{tab:b}
\end{table}

\section{CONCLUSION}

We have shown that the destroyed symmetry of the electrons in a quantum dot
can be recovered by adjusting the magnetic field and that there is an
infinite set of exact analytical solutions of the Schr\"{o}dinger equation
for two electrons in an external homogeneous magnetic field. It is easy to
see that the confinement region of the electrons in the dot can be expressed
as%
\begin{equation}
\ell =\left( \frac{\hbar }{m^{\ast }\omega }\right) ^{1/2}.  \label{e21}
\end{equation}%
This may be interpreted as follows: The size of the QD depends on the
quantum number of the angular momentum and the external magnetic field. When
the magnetic field increases, the dot size will decrease, and the electrons
have higher angular momentum.

Finally, we have mention that the method given in this article can be
extended in various directions. The algebra can be modified to investigate
the hidden symmetries of the interacting electrons in quantum dots\cite{kim}%
. As it is known that spin-related phenomena are the key ingredient in the
emerging field of spintronics\cite{wolf}, the Hamiltonian representing the
Rashba and Dresselhaus spin-orbit coupling can be treated by using the $%
osp(2,1)$ or the $osp(2,2)$ algebra within the framework of quasi-exact
solvability.

\end{document}